\documentclass[showpacs,twocolumn,preprintnumbers,amsmath,amssymb,prl,epsf,superscriptaddress]{revtex4-1}
\usepackage{graphicx}
\usepackage{nicefrac}
\usepackage[squaren]{SIunits}
\usepackage{color}

\begin{document}
\title{Giant Narrowband Twin-Beam Generation along the Pump Energy Propagation}
\author{Angela M.~P\'erez}
\affiliation{Max Planck Institute for the Science of Light,
G\"unther-Scharowsky-Stra\ss{}e 1/Bau 24, 91058 Erlangen, Germany}
\affiliation{University of Erlangen-N\"urnberg, Staudtstrasse 7/B2, 91058 Erlangen, Germany}
\author{Kirill Yu.~Spasibko}
\affiliation{Max Planck Institute for the Science of Light,
G\"unther-Scharowsky-Stra\ss{}e 1/Bau 24, 91058 Erlangen, Germany}
\affiliation{University of Erlangen-N\"urnberg, Staudtstrasse 7/B2, 91058 Erlangen, Germany}
\affiliation{Department of Physics, M.V.Lomonosov Moscow State University, \\ Leninskie Gory, 119991 Moscow,
Russia}
\author{Polina R.~Sharapova}
\affiliation{Department of Physics, M.V.Lomonosov Moscow State University, \\ Leninskie Gory, 119991 Moscow,
Russia}
\author{Olga V.~Tikhonova}
\affiliation{Department of Physics, M.V.Lomonosov Moscow State University, \\ Leninskie Gory, 119991 Moscow,
Russia}
\author{Gerd Leuchs}
\affiliation{Max Planck Institute for the Science of Light, G\"unther-Scharowsky-Stra\ss{}e 1/Bau 24, 91058
Erlangen, Germany} \affiliation{University of Erlangen-N\"urnberg, Staudtstrasse 7/B2, 91058 Erlangen, Germany}
\author{Maria~V.~Chekhova}
\affiliation{Max Planck Institute for the Science of Light, G\"unther-Scharowsky-Stra\ss{}e 1/Bau 24, 91058
Erlangen, Germany} \affiliation{Department of Physics, M.V.Lomonosov Moscow State University, \\ Leninskie Gory,
119991 Moscow, Russia}
\affiliation{University of Erlangen-N\"urnberg, Staudtstrasse 7/B2, 91058 Erlangen, Germany}

\vspace{-10mm}

\vspace{5mm}
\maketitle \narrowtext

\textbf{Walk-off effects, originating from the difference between the group and phase velocities, limit the efficiency of nonlinear optical interactions~\cite{Boyd,Agrawal}. While transverse walk-off can be eliminated by proper medium engineering, longitudinal walk-off is harder to avoid. In particular, ultrafast twin-beam generation via pulsed parametric down-conversion (PDC) and four-wave mixing (FWM) is only possible in short crystals~\cite{Bouwmeester} or fibres~\cite{Fan,Smith} or in double-path schemes~\cite{Eisenberg}. Here we show that in high-gain PDC~\cite{Lugiato,Bondani,Genovese,Iskhakov}, one can overcome the destructive role of both effects and even turn them into useful tools for shaping the emission. In our experiment, one of the twin beams is emitted along the pump Poynting vector or its group velocity matches that of the pump. The result is dramatically enhanced generation of both twin beams, with the simultaneous narrowing of angular and frequency spectrum. The effect will enable efficient generation of ultrafast twin photons and beams in cavities~\cite{cavity}, waveguides~\cite{WG}, and whispering-gallery mode resonators~\cite{WGM}.}

Walk-off effects are caused by the difference between the velocities and directions of energy and phase propagation. In particular, transverse (spatial) walk-off is due to the fact that the Poynting vector is, in the general case, noncollinear to the wavevector. Similarly, longitudinal (temporal) walk-off appears because the energy of a pulse travels at a different speed than its phase. Since in nonlinear optical processes it is \textit{phase} matching that determines the frequency and angular spectrum, interacting pulses are, in the general case, \textit{group mismatched} and separate in both space and time in the course of propagation. This limits the length of nonlinear interactions for short pulses and focused beams.

This is definitely the case for high-gain parametric down-conversion (PDC), widely used for the generation of bright squeezed vacuum~\cite{Lugiato,Bondani,Genovese,Iskhakov,Agafonov}. This nonclassical state of light has interesting applications in quantum imaging~\cite{Lugiato,Brida} and sensing~\cite{illumination}, phase super-resolution~\cite{Agarwal}, macroscopic entanglement~\cite{nonseparability}, enhanced nonlinear interactions and many other fields.
In high-gain PDC, down-converted radiation is exponentially amplified provided that it overlaps in space and time with the pump pulse. This makes certain angles and frequencies of emission distinguished and the emission into the corresponding modes enhanced.
\subsection{High-gain PDC along the pump Poynting vector}
\begin{figure}[h]
\begin{center}
\includegraphics[width=0.35\textwidth]{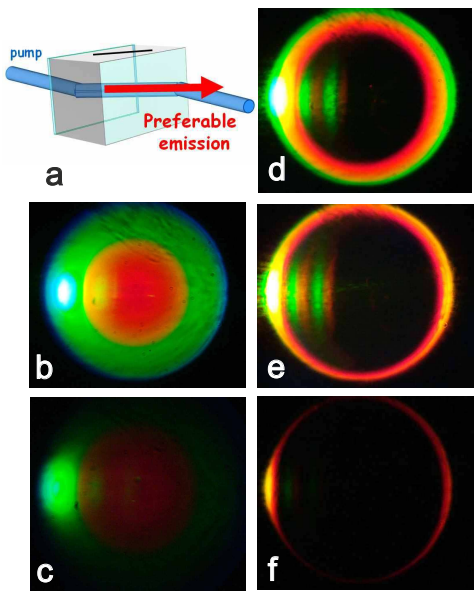}
\caption{Beamlike PDC along the pump spatial walk-off. The highest gain is achieved for emission along the pump Poynting vector, which inside the crystal is non-collinear to the wavevector (a). The optic axis direction is schematically shown on top of the crystal. Snapshots with a photographic camera (b-f) show the spectra of high-gain PDC at different crystal orientations. Additional stripes in d,e are due to the reflections in the crystal. Emission at the walkoff angle and in the phase-matched direction is enhanced. This occurs, at different crystal orientations, for different colours (d-f). A snapshot with a neutral-density filter (c) shows colours without the saturation.}\label{general}
\end{center}
\end{figure}

For PDC in an anisotropic crystal, if the pump is polarized extraordinarily and the down-converted light ordinarily, the emission will be mainly along the pump Poynting vector (Fig.~\ref{general}a) as well as along its matching direction. The latter originates from the fact that every signal photon emitted in the walk-off direction has its idler counterpart, and vice versa. This leads to a high-intensity beamlike PDC.

\begin{figure}[h]
\begin{center}
\includegraphics[width=0.5\textwidth]{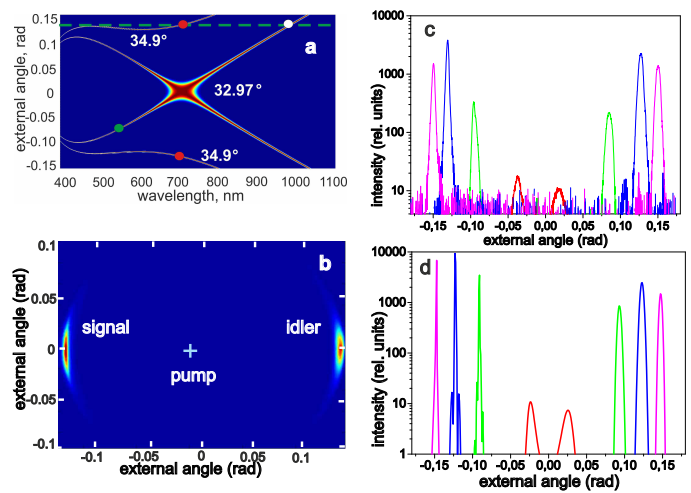}
\caption{Tuning curves of type-I PDC from $355$ nm pump (a) show that at orientation $32.97^\circ$, only the infrared beam (white dot) propagates along the Poynting vector (green dashed line), but its twin is in the green spectral range (green dot). With the crystal at $34.9^\circ$, the emission at the degenerate wavelength $710$ nm is at the walk-off angle (red dots), and the 2D intensity distribution obtained by the CCD camera (b) shows hugely enhanced low-divergence twin beams. The rest of the PDC ring is much weaker and not seen in the linear scale. At other crystal orientations, emission at $710$ nm is much less intense, which is shown by logarithmic-scale 1D angular intensity distributions measured (c) and calculated (d).}\label{intensity}
\end{center}
\end{figure}

In our experiment, we focused the Gaussian pump beam (frequency-tripled YAG:Nd laser with the wavelength 355 nm, pulse duration 18 ps and repetition rate 1 kHz) into a 5 mm BBO crystal by means of a lens with the focal length $500$ mm, which resulted in a waist with full width at half maximum (FWHM) of $60\,\mu$m. The crystal was oriented to provide frequency-degenerate collinear type-I phase matching. After the crystal, the pump was cut off by a dichroic mirror. The angular spectra of PDC were first captured using a digital photographic camera (Fig.~\ref{general}b). We see beamlike strongly enhanced emission in the green spectral range.  The color is modified because of the camera saturation, which is clear from a snapshot with a neutral-density filter (Fig.~\ref{general}c). The green beam is only one of the `twins'; moreover, it is the other beam that propagates along the Poynting vector ($4^\circ$ to the collinear direction), but it belongs to the IR range and is therefore not visible (Fig.~\ref{intensity}a).

By tilting the crystal, one can change the phase matching, so that the beamlike emission is observed in other spectral ranges (Figs.\ref{general}d,e,f). As the wavelength of the enhanced beam approaches the visible range, both left and right parts of the ring become pronounced (Fig.~\ref{general}f).

To demonstrate both twin beams emitted at the degenerate wavelength, we installed a bandpass filter (bandwidth 10 nm centered at 710 nm) after the crystal and tilted the crystal to the orientation $34.9^\circ$. The resulting angular distribution was recorded by a CCD camera (Fig.~\ref{intensity}b). One can hardly see the whole ring of emission at 710 nm (in fact it is seen with a much larger exposure time) but there is a strong peak in the direction corresponding to the walk-off angle in the crystal, as well as in the direction symmetrical with respect to the pump.

At other orientations of the crystal, the emission at the degenerate wavelength is much weaker, as shown in Fig.~\ref{intensity}c), which presents 1D cuts of the obtained 2D spectra. In agreement with the theory (Fig.~\ref{intensity}d), the PDC peak observed in the walk-off direction exceeds the near-collinear emission by more than two orders of magnitude.

As seen in Fig.~\ref{intensity}b, the angular divergence of both beams is very small. This is because amplification occurs within the angle $a/L$, where $a$ is the pump beam waist and $L$ is the crystal length. Because the pump divergence, $\lambda_p/a$, is on the same order of magnitude, the beams are nearly single-mode. This effect is similar to the generation of spatially single-mode PDC from two separated crystals~\cite{separation}. Indeed, in experiment we measured the effective number of spatial modes $m=1.4$ at the maximal parametric gain of $15$, corresponding to the pump power of $30$ mW~\cite{sup}.

Note that at low-gain PDC no such beamlike emission can be observed, only asymmetry in the angular spectrum~\cite{Fedorov_PRL}. The effect at high gain is known and widely used in parametric amplification~\cite{Gale} but, to the best of our knowledge, has been never observed in twin-beam generation. A very special feature of the current configuration is that high directionality of radiation is achieved without using a cavity, and that the wavelength of the enhanced radiation is tunable within a broad range. The idler beam can serve as a source of tunable broadband diffraction-limited infrared radiation, especially convenient because its copy is maintained in the visible-range signal beam.

\subsection{High-gain PDC with group-velocity matching}
While spatial walk-off can be, in principle, eliminated by using noncritical phase matching, double-crystal configuration~\cite{Grangier,anisotropy1} or periodically poled materials, temporal walk-off is inevitable whenever PDC or FWM is pumped by ultrashort pulses. Due to the group velocity difference between the pump and the twin-beam radiation, the pump pulse is delayed from the signal and idler pulses in the course of propagation through the nonlinear material, and only short crystals or fibres can be used for coherent generation of twin beams. For instance, in the common case of collinear frequency-degenerate type-I PDC generated in a BBO crystal from the frequency-doubled Ti-sapphire laser ($400$ nm), a 180 fs delay emerges in a 1 mm crystal. However, at some other wavelength of the signal radiation its group velocity can be equal to that of the pump. The phasematching can then be fulfilled by using noncollinear emission. The same situation can be realized for type-II PDC. Note that only one pulse, signal or idler, has to propagate together with the pump pulse. Its twin will be amplified as well, similarly to the spatial walk-off case. Hence, giant amplification of both twin beams should be expected.

\begin{figure}[h]
\begin{center}
\includegraphics[width=0.47\textwidth]{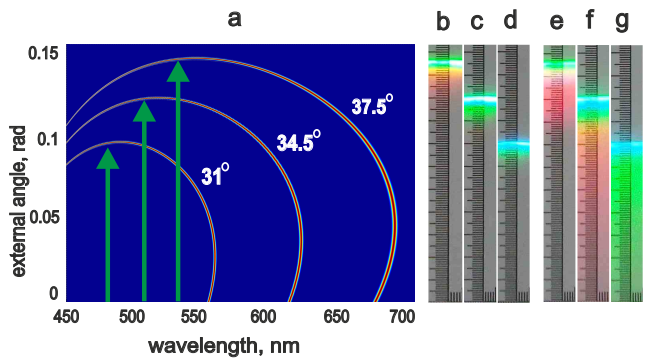}
\caption{Group matching for type-II PDC in a BBO crystal pumped by $1.2$ ps pulses at $400$ nm. Tuning curves for the ordinary down-converted radiation are shown for crystal orientations $31^\circ$, $34.5^\circ$, and $37.5^\circ$ (a). For each orientation, at some wavelength the group velocity of the ordinary photon matches that of the pump (green arrows). The PDC ring at this wavelength is hugely enhanced if the crystal is long enough (we used four 5 mm BBO crystals stacked together). Panels b-d show snapshots of the intensity distributions on a screen placed in the far field, with the angular scale and the crystal orientations the same as in panel a. For comparison, spectra of a single 5 mm crystal with the same orientations show much broader emission spectra (e,f,g).}\label{spectra}
\end{center}
\end{figure}

In our experiment, we used type-II PDC in a BBO crystal pumped at $400$ nm. Tuning curves of the `ordinary' beam for positive angles are shown in Fig.~\ref{spectra}a for the orientations $31^\circ,34.5^\circ$ and $37.5^\circ$. Green arrows mark the positions of group-velocity matching. One should expect considerable amplification of emission at these wavelengths in the case of a long crystal and high gain.

PDC was pumped by $1.2$ ps pulses of frequency-doubled Mai Tai/Spitfire radiation with the repetition rate $5$ kHz and the energy per pulse up to $0.1$ mJ. To demonstrate the temporal walk-off, we reduced the spatial walk-off by focusing the pump beam only in the plane orthogonal to the principal one by means of cylindrical optics. In the principal plane, the beam FWHM was $3.4$ mm but in the orthogonal plane, 130 $\mu$ or 680 $\mu$ depending on the configuration. The spectra obtained from a single $5$ mm crystal were compared with the ones of four $5$ mm crystals placed one after another at a minimal distance $3$ mm.
The effect is easily observable on a screen (Fig.~\ref{spectra}). For PDC generated by 20 mm of BBO, there is huge amplification within a relatively narrow spectral range (Fig.~\ref{spectra} b-d), while in the case of a 5 mm crystal the spectrum is more uniform over frequency (Fig.~\ref{spectra} e-g). The angle at which amplification
occurs is different depending on the orientation; this clearly shows that it is not related to the spatial walk-off.

\begin{figure}[h]
\begin{center}
\includegraphics[width=0.3\textwidth]{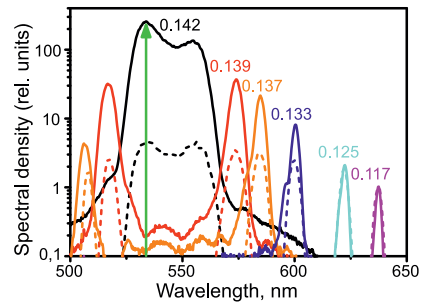}
\caption{Frequency spectra of PDC recorded at external angles $0.142$, $0.139$, $0.136$, $0.128$,  and $0.117$ rad for one (dashed line) and four (solid line) BBO crystals of length $5$ mm oriented at $37.5^\circ$. In $20$ mm of BBO crystal, the emission at the group-matched wavelength $533$ nm (shown by green arrow) is enhanced $250$ times compared to the wavelength $636.5$ nm, where there is no group-velocity matching. In a $5$ mm crystal, the enhancement is not so dramatic.}\label{spectrum}
\end{center}
\end{figure}

For the quantitative characterization of the effect, we recorded the frequency spectra at different angles, with the crystals oriented at $37.5^\circ$. According to the tuning curve (Fig.~\ref{spectra}a), each angle corresponds to a peak at a different wavelength. For PDC from $20$ mm BBO, the peak at the group-velocity-matching wavelength $533.5$ nm exceeds the one at $636.5$ nm $250$ times (Fig.~\ref{spectrum}). Because this difference is influenced by many frequency-dependent factors~\cite{Klyshko,Spasibko}, we also recorded spectra for a $5$ mm crystal. Although the temporal walk-off is not negligible also in this case, the spectral distribution is much more uniform. Note that in both cases, the parametric gain was made the same ($G=8.6$) at the wavelength $636.5$ nm, by stronger focusing the pump in the case of a $5$ mm crystal. The focusing was performed by a cylindrical lens (focal length $500$ mm) or a cylindrical telescope (30:2.5) in order to avoid the transverse walk-off.

This effect can be especially useful for generating twin beams in whispering-gallery mode resonators~\cite{WGM} where, due to the large quality factor, nonlinear interaction occurs on a very large distance. This requires pump beams with large coherence length; however, if the group velocity of down-converted radiation at a certain wavelength coincides with that of the pump, strong twin-beam generation can be realized even from short pulses. Under certain conditions, this radiation will be single-mode.

The effect will have interesting consequences for low-gain PDC. Indeed, the condition of temporal walk-off compensation, in combination with a large crystal length, will provide the absence of frequency correlations for femtosecond-pulsed PDC~\cite{Grice,Mosley}. At the same time, signal and idler photons will have very different spectral widths and, consequently, different pulse durations.

In periodically poled crystals or waveguides, phase matching can be satisfied also in the collinear regime by a proper poling. Near resonance, one can have the pump group velocity very different from the signal/idler group velocity. This could provide additional possibilities for engineering the spectrum.

\subsection{Conclusion} We have shown that, in contrast to the common opinion, spatial and temporal walk-off is not necessarily degrading the generation of twin beams. When properly used, it can lead to the generation of bright, tunable, and diffraction-limited twin-beam radiation. So far we have demonstrated the amplification of down-converted radiation by more than two orders of magnitude under the condition that the signal or idler pulses propagate in the same direction or with the same velocity as the pump pulse energy. More dramatic amplification can be expected for PDC or FWM in long nonlinear media (fibres, waveguides) or resonators.

\section{Acknowledgements}
We acknowledge the financial support of the European Union under project BRISQ2 (FP7-ICT) and of the Russian Foundation for Basic Research, grants 14-02-31030, 14-02-00389, and 14-02-31084.

\section{Methods}
\subsection{Calculation of high-gain PDC spectra}
Our model~\cite{Bloch,separation} is based on the Bloch-Messiah reduction.
Because the anisotropy manifests itself significantly only in the principal plane of the crystal, the Hamiltonian can be calculated using one Cartesian dimension~\cite{Bloch,anisotropy1}.
By passing to collective photon creation operators $A_{n}^\dagger, B_{n}^\dagger$ corresponding to the Schmidt modes $u_n(\vartheta_s),v_n(\vartheta_i)$, where $\vartheta_s,\vartheta_i$ are the signal and idler angles inside the crystal, we diagonalize the Hamiltonian as
\begin{equation}
H=i\hbar\Gamma \sum_{n} \sqrt{\lambda_{n}}(A_{n}^\dagger B_{n}^\dagger-A_{n} B_{n}),
\label{Hams}
\end{equation}
where $\lambda_{n}$ are the Schmidt eigenvalues. Further, we write time-depending differential equations for the new operators in the Heisenberg representation,
and find their solutions given by the Bogolyubov transformations~\cite{sup}.
The equations for the plane-wave operators can be obtained and solved analytically in the Heisenberg picture using the expression for the Schmidt operators. Then the mean photon number in the signal beam can be calculated by averaging over the vacuum state and is
\begin{equation}
\langle N_s \rangle=\sum_n \vert u_n(\vartheta_s) \vert^2( \sinh[\sqrt{\lambda_n}G])^2,
\label{Ns}
\end{equation}
and similarly for the idler beam. Here, $G\equiv\int2\Gamma dt$ is the parametric gain. Thus, the total intensity distribution results from incoherent contributions of Schmidt modes with new weights that differ dramatically from the initial Schmidt eigenvalues $\lambda_n$ and strongly depend on the parametric gain. The renormalized new weights of different Schmidt modes are
\begin{equation}
\Lambda_n=\frac{(\sinh[G\sqrt{\lambda_{n}}])^2}{\sum_n (\sinh[G\sqrt{\lambda_{n}}])^2}.
\label{lambda}
\end{equation}
\subsection{Measurement of angular spectra}
For recording the angular spectra, we placed an aspheric lens with the focal length $f=26$ mm and numerical aperture $0.5$ at a distance about $3$ cm from the crystal~\cite{sup} and a CCD camera into the focal plane of the lens. The 2D spatial intensity distributions recorded by the camera corresponded then to the angular distributions of PDC radiation, with the external (outside of the crystal) angle $\theta$ given by the coordinate $x$ as $\theta=\arctan(x/f)$. The relation between the internal angles $\vartheta_{i,s}$ and the external ones $\theta_{i,s}$ is given by Snell's law, $\sin\theta_{i,s}/\sin\vartheta_{i,s}=n_{i,s}$, with $n_{i,s}$ being the refractive indices for the idler and signal radiation.  The one-dimensional spectra were obtained as cross-sections of the 2D distributions.
\subsection{Measurement of frequency spectra}
The frequency spectra were recorded by means of an Ocean Optics HR4000 spectrometer connected to a $400\mu$m multimode fibre placed at $84$ cm from the crystals.
\subsection{Measurement of the number of modes}
The number of spatial modes was determined by collecting one of the twin beams into the entrance slit of a monochromator with the resolution $0.1$ nm and then, after collimating the output beam, sending it into a Hanbury Brown-Twiss interferometer~\cite{sup} for measuring the normalized second-order correlation function $g^{(2)}$. The number of modes $m$ can be found from the relation $g^{(2)}=1+1/m$~\cite{separation}.

\end{document}